%% file: spdf_oscillation.tex
\documentclass[aps,prl,twocolumn]{revtex4-1}

\usepackage[usenames,dvipsnames,svgnames,table]{xcolor}
\usepackage{graphicx}
\usepackage[dvipsnames]{xcolor}
\usepackage{amsmath}
\usepackage{blindtext}
\usepackage{printlen}
\usepackage{hyperref}
\usepackage{braket}
\usepackage[disable]{todonotes}
\usepackage[pass]{geometry}
\uselengthunit{in}
\usepackage{soul}

\usepackage{CJKutf8} 

\newcommand{\comment}[2][]{\todo[color=red!100!green!33, #1]{#2}}
\definecolor{myellow}{rgb}{1., 1., 0.6}

\begin{document}
\begin{CJK*}{UTF8}{gbsn}

\title{Tunable Single-Ion Anisotropy in Spin-1 Models Realized with Ultracold Atoms}
            
\author{Woo Chang Chung}
\thanks{These authors contributed equally to this work.}
\author{Julius de Hond}
\thanks{These authors contributed equally to this work.}
\author{\CJKfamily{gbsn}Jinggang Xiang (项晶罡)}
\author{Enid Cruz-Col\'{o}n}
\author{Wolfgang Ketterle}
\affiliation{Research Laboratory of Electronics, MIT-Harvard Center for Ultracold Atoms, Department of Physics, Massachusetts Institute of Technology, Cambridge, Massachusetts 02139, USA}

\date{\today}

\begin{abstract}
Mott insulator plateaus in optical lattices are a versatile platform to study spin physics. Using sites occupied by two bosons with an internal degree of freedom, we realize a uniaxial single-ion anisotropy term proportional to $(S^z)^2$, which plays an important role in stabilizing magnetism for low-dimensional magnetic materials. Here we explore non-equilibrium spin dynamics and observe a resonant effect in the spin anisotropy as a function of lattice depth when exchange coupling and on-site anisotropy are similar. Our results are supported by many-body numerical simulations and are captured by the analytical solution of a two-site model.
\end{abstract}

\maketitle
\end{CJK*}

Mott insulators of ultracold atoms in optical lattices comprise a widely used platform for quantum simulations of many-body physics \cite{bloch2008many}.  Since the motion of atoms is frozen out, the focus is on magnetic ordering and spin dynamics in a system with different (pseudo-)spin states.  As suggested in 2003, Mott insulators with two-state atoms realize quantum spin models with tunable exchange interactions and magnetic anisotropies \cite{duan2003controlling,kuklov2003counterflow}. Experimental achievements for spin-1/2 systems include the observation of antiferromagnetic ordering of fermions \cite{Mazurenko2017}  and the study of spin transport in a Heisenberg spin model with tunable anisotropy of the spin-exchange couplings \cite{jepsen2020}.  Spin dynamics for $S > 1$ has also been investigated  \cite{dePaz2013}. 

However, all studies thus far have exclusively addressed spin systems with occupations of one atom per site. This limits spin Hamiltonians to spin-exchange terms between different sites $i, j$ proportional to $\sum_{\langle ij \rangle} S_{i}^{k}S_{j}^{k}$ (where $k \in \left\{x, y, z\right\}$) and to Zeeman couplings to effective magnetic fields, proportional to $\sum_{i} S_{i}^{z}$.  For Mott insulators with two or more atoms per site, the Hubbard model has direct on-site interactions which can give rise to a nonlinear term $D \sum_{i} (S_{i}^{z})^2$, where $D$ is the so-called single-ion anisotropy constant. $(S^z)^2$ terms, which are present for $S\geq 1$ only, are important for establishing non-trivial correlations, such as in spin squeezing~\cite{KitagawaUeda}. In spin-1 models, they can lead to a qualitatively new magnetic phase diagram \cite{Li2011, Li2016}. For example, for ferromagnetic spin-1 Heisenberg models, the single-ion anisotropy gives rise to a gapped spin state (the ``spin Mott insulator'') that can be used as an initial low-entropy  state for an adiabatic ramp toward a highly-correlated gapless spin state (the XY ferromagnet) \cite{altman2003phase,Schachenmayer2015}. For antiferromagnetic systems in one dimension, the single-ion anisotropy leads to a quantum phase transition between a topologically trivial phase and a nontrivial phase as predicted by Haldane \cite{Haldane1983,Haldane1983PRL, Haldane2017}. Magnetic properties of many materials crucially depend on crystal field anisotropies which break rotational symmetry and can stabilize ferromagnetism in two-dimensional materials by avoiding the Mermin--Wagner theorem which forbids long-range order for continuous symmetries \cite{merminwagner1966, Strecka2008}. The interest in spin-1 systems is demonstrated by various studies on different platforms \cite{Renard1987,Chauhan2020,Senko2015}.

\begin{figure}
\centering
\includegraphics[width=\columnwidth]{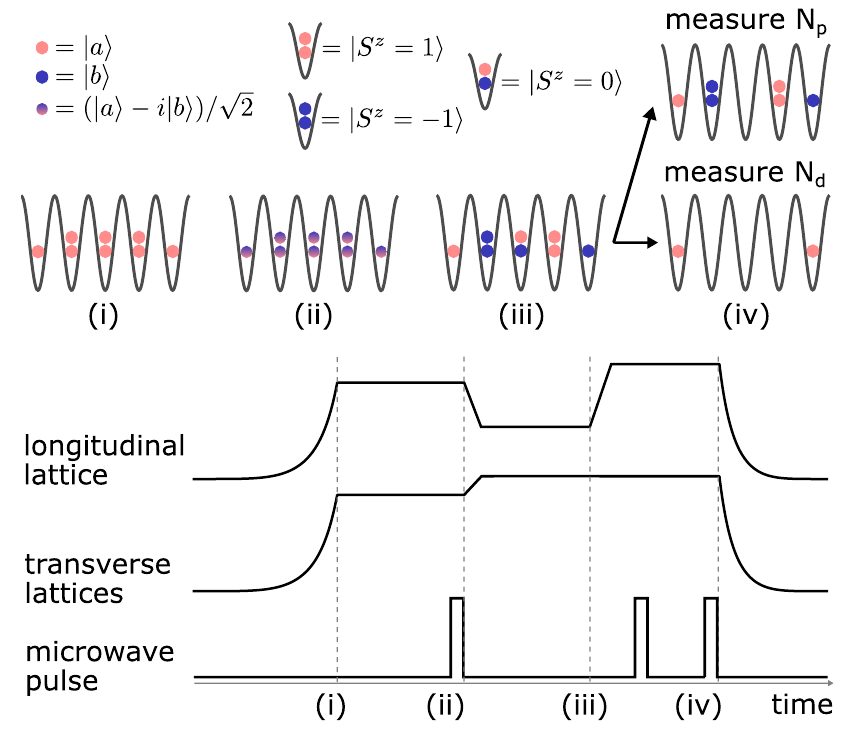}
\caption{Experimental sequence for the measurement of spin alignment and doublon fractions. (i) The lattices are ramped up to initialize a single-component Mott insulator with a maximal site occupancy of two. (ii) Microwave pulses prepare a superposition of two hyperfine states $\left(|a\rangle -i |b\rangle\right)/\sqrt{2}$. Ramping down the longitudinal lattice initiates spin exchange dynamics. (iii) Ramping up the lattices stops the exchange dynamics.  Microwave pulses transfer the two components to a pair of states with a Feshbach resonance. (iv) Either $|a b\rangle$ doublons or all doublons are removed with the help of Feshbach-enhanced inelastic losses. Remaining atoms are transferred back to the $F=1$ hyperfine states and are counted via absorption imaging to measure $N_p$ or $N_d$.} 
\label{fig:exp_procedure}
\end{figure}

In this Letter, we use cold atoms in optical lattices to implement a spin-1 Heisenberg Hamiltonian using a Mott insulator of doubly occupied sites and demonstrate unique dynamical features of the single-ion anisotropy. For spin-exchange interactions studied thus far in optical lattices, the only time scale for dynamics is second-order tunneling (i.e.\ superexchange) which monotonically slows down for deeper lattices.  In contrast, as we show here, the single-ion anisotropy introduces a new time scale, and we find a dynamical behavior which is faster in deeper lattices, due to a resonance effect when the energies of superexchange and single-ion anisotropy are comparable. 

We present a protocol to directly measure the anisotropy in the spin distribution and find pronounced transient behaviour of this quantity when the resonance condition is met. Transients change sign along with the the single-ion anisotropy. We find good agreement with theoretical simulations, and explain the most salient features using a two-site model with an exact solution.

In the Mott insulator regime the optical lattices are sufficiently deep such that first-order tunneling is suppressed, and exchange processes are only possible via second-order tunneling. For two atoms per site, the Bose--Hubbard Hamiltonian is approximated by an effective spin Hamiltonian
\begin{eqnarray}\label{eq:spin-1-Hamiltonian}
    H &=&-J\sum_{\langle ij \rangle}\mathbf{S}_i \cdot \mathbf{S}_j+D\sum_{i}(S_{i}^z)^2 - B\sum_{i}S_{i}^z,
\end{eqnarray}
where $\mathbf{S}_i$ are spin-1 operators, $\langle ij \rangle$ are pairs of nearest-neighboring sites, $J$ is the exchange constant, $D$ is the uniaxial single-ion anisotropy constant, and $B$ is a fictitious magnetic bias field. The spin-1 operators are related to the boson creation/annihilation operators via $S_{i}^{z} = (a_i^\dagger a_i -b_i^\dagger b_i)/2$, $S_{i}^{+} = a_i^{\dagger}b_i$, $S_{i}^{-} = b_i^{\dagger}a_{i}$ under the constraint $a_{i}^{\dagger}a_{i} + b_{i}^\dagger b_{i} = 2$, where $a_{i}$ and $b_{i}$ are boson annihilation operators at site $i$ for state $a$ and state $b$ respectively. In terms of the tunneling amplitude $t$ and interaction energies $U_{\sigma\sigma'}$: $J = 4 t^2 / U_{ab}$ and $D = (U_{aa} + U_{bb})/2 - U_{ab}$, where $U_{\sigma\sigma'}$ represents the on-site interaction energy between atoms in two states $\sigma,\sigma' \in \{a,b\}$. The term proportional to $B$ can be dropped if the total longitudinal magnetization $\sum_{i} S_{i}^{z}$ is constant, as it is in the experiment.  

For the species studied here, $^{87}$Rb, all $U_{\sigma\sigma'}$ differ by less than 1\%, and therefore all spin exchange couplings are almost equal resulting in isotropic spin Hamiltonians for site occupancy $\nu=1$. However, for $\nu=2$, we can tune the relevant anisotropy parameter $D/J$ over a large range of values, because $J$ decreases exponentially with lattice depth, while $D$---a differential on-site energy---slowly increases.

The experimental sequence begins by preparing a Bose--Einstein condensate (BEC) of $^{87}$Rb atoms in the $|F=1,m_F = -1\rangle$ hyperfine state inside a crossed optical dipole trap. It proceeds by loading the BEC into a deep three-dimensional optical lattice formed by retro-reflected lasers with wavelengths of $\lambda = 1064~\mathrm{nm}$. The lattices are ramped to final depths of $30\,E_R$ in 250 ms, where $E_R = h^2/\left(2 m \lambda^2 \right)$ is the recoil energy for atomic mass $m$. Experimental parameters are chosen to maximize the size of the  $\nu=2$ Mott-insulator plateau without significant population of sites with  $\nu=3$ [see Fig.~\ref{fig:exp_procedure}(i) and the Supplemental Material].

To allow for spin dynamics, all atoms are rotated into an equal superposition of two hyperfine states $(|a\rangle-i|b\rangle)/\sqrt{2}$ using a combination of microwave pulses (see the Supplemental Material). This initial state is a simple product state. Negative and positive values of $D$ are realized with the pairs $|a\rangle = |1,-1\rangle$, $|b\rangle = |1,1\rangle$, and $|a\rangle = |1,-1\rangle$, $|b\rangle = |1,0\rangle$, respectively (see the Supplemental Material). The spin exchange dynamics in one-dimensional chains is initiated by a 3-ms quench, during which we ramp down the longitudinal lattice to a variable depth, while the transverse lattices are ramped up to $35\,E_R$ [Fig.~\ref{fig:exp_procedure}(ii)].  After a variable evolution time, the final spin configuration is ``frozen in'' by ramping the longitudinal lattice to $35\,E_R$ as well [Fig.~\ref{fig:exp_procedure}(iii)].

Our observable for the anisotropy in the spin distribution is the longitudinal spin alignment $A = S(S+1) -3\langle (S^z)^2\rangle$, measured in the $\nu=2$ plateau. $\langle (S^{z})^2 \rangle = \sum_{i=1}^{N} \langle (S_{i}^{z})^2\rangle/N$ is the average on-site longitudinal spin correlation. $A$ is defined to be zero for a random distribution of spins.  Since $S^z= 1,  0, -1$ for the $\ket{aa}$, $\ket{ab}$ and $\ket{bb}$ doublons, respectively, $A$ can be obtained by measuring the relative abundance of the different doublons. Specifically, we refer to the fraction of $\ket{ab}$ doublons as the ``spin-paired doublon fraction'' $f$.  Since $\langle (S^{z})^2 \rangle = 1 - f$, we obtain $A = 3f - 1$.  The doublon statistics can be measured by selectively introducing a fast loss process that targets a specific type of doublon, and by comparing the remaining total numbers of atoms, which are measured via absorption imaging. Specifically, if $N_{a}$ is the average total atom number in the whole cloud, $N_{p}$ the average number of remaining atoms after removing $\ket{ab}$ doublons, and $N_{d}$ the average number of remaining atoms after removing all doublons, then $f= (N_a - N_p)/(N_a-N_d)$ [Fig.~\ref{fig:exp_procedure}(iv)]. 
Fast losses of doublons are induced by transferring the atoms to hyperfine states for which inelastic two-body loss is enhanced near two narrow Feshbach resonances around a magnetic field of $9~\mathrm{G}$~\cite{kaufman2009radio} (also see the Supplemental Material).  Since $f$ and $A$ are obtained from the ratio of differences in atom numbers, good atom number stability in the experiment (the deviation from mean being typically $<4$\,\%) was crucial to measure $A$ with sufficiently small uncertainties.

\begin{figure}
    \centering
    \includegraphics[width=\columnwidth]{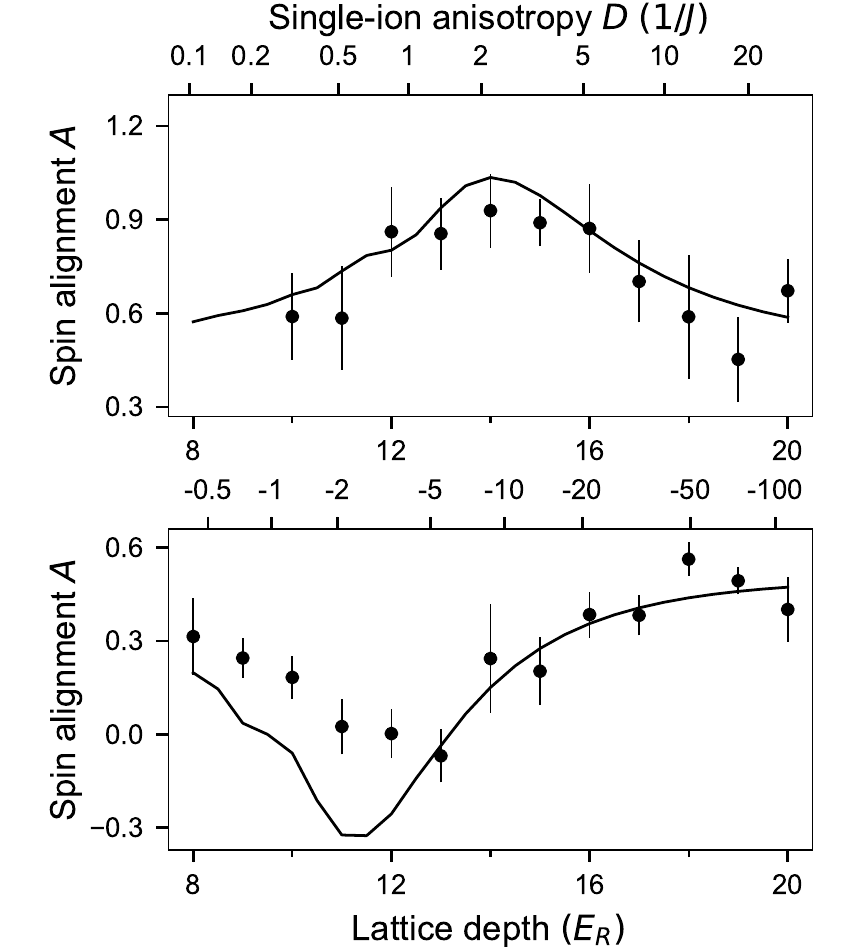}
    \caption{Transient enhancement and reduction of the spin alignment $A$ by coherent spin dynamics. The change in $A$ is strongest when $\left| D/J \right| \sim 2$. Measurements were done for both  positive (top) and negative (bottom) values of  $D/J$. The atoms were held for $70~\mathrm{ms}$ and $25~\mathrm{ms}$, respectively (also see Fig.~\ref{fig:hold_time_scan}). The top axis in both figures indicates the $D/J$ ratio. Solid lines are the results of MPS-TEBD calculations. The error bars represent the standard error of the mean for $A$, obtained by error propagation after averaging three measurements for each of $N_a$, $N_p$, and $N_d$. For the lowest lattice depths, the spin model may not fully represent the Bose--Hubbard model.}
    \label{fig:lattice-depth-scan}
\end{figure}

For the initial state, $f=1/2$ and $A= 1/2$. Over times that are long compared to spin exchange time scale $\hbar/J$, heating processes drive the system towards thermal equilibrium with $A=0$. At short times, coherent spin dynamics is observed: If $D$ is negative, the $\ket{aa}$  and $\ket{bb}$ doublons are energetically favorable, and we expect $f$ and $A$ to decrease. If $D$ is positive, the $\ket{ab}$ doublons are favorable and we expect $f$ and $A$ to increase. If $D$ is zero, the system is described by an isotropic spin-1 Heisenberg Hamiltonian of which the initial state is an eigenstate. By fixing the hold time and scanning the value of the lattice depth for the spin chains, we can monitor the impact of $D/J$ on the dynamical change in $A$. For positive (negative) $D$, we chose a hold time of $70~\mathrm{ms}$ ($25~\mathrm{ms}$). These hold times are chosen to be comparable to $\hbar/J$ when $|D/J| \sim 1$ (see the Supplemental Material).

Figure~\ref{fig:lattice-depth-scan} shows that for $|D/J| \ll 1$ or $|D/J| \gg 1$, $A$ stays near its initial value of $1/2$.  However, when $D/J \sim~2$, which corresponds to a longitudinal lattice depth of $14\, E_R$ ($11\,E_R$) for positive $D$ (negative $D$), we see that $A$ reaches a maximum (minimum). This non-monotonic change of $A$ with lattice depth is indicative of the interplay between spin exchange and single-ion anisotropy.  In addition, we observe that the change in $A$ is smaller for positive $D$ than for negative $D$. 

\begin{figure}
    \centering
    \includegraphics[width=\columnwidth]{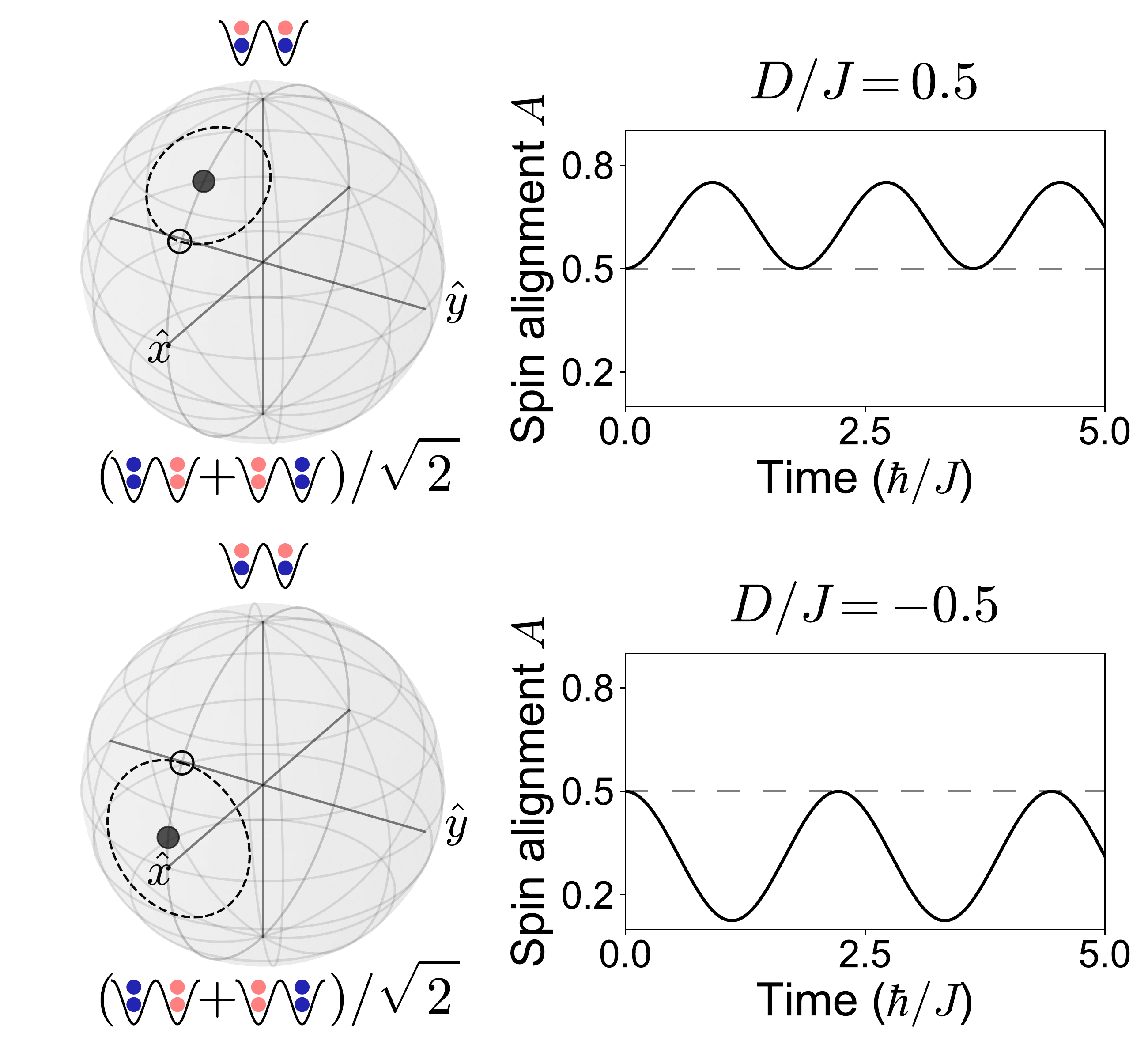}
    \caption{Coherent spin oscillations in a two-site model. While the full basis contains nine states, the oscillations in the spin alignment $A$ involve only a $2\times 2$ block of the Hamiltonian. This allows us to illustrate the dynamics on Bloch spheres (left), where the initial state is represented by the open circle. If $J = 0$ the effective magnetic field points along $\hat{z}$, and the purely azimuthal precession will not change $A$. If $J > 0$ the effective magnetic field is tilted, resulting in a precession along the dashed circle which is observed as an oscillation in $A$ (right). The frequency of the oscillation, in units of $J/\hbar$, is given by $\Omega = \sqrt{9 + 4D/J + 4\left( D/J \right)^2}$, and its amplitude is $2\left(D/J \right)/\left[ 9 + 4D/J + 4\left(D/J \right)^2 \right]$. This shows that the direction of oscillation depends on the sign of $D/J$ (compare top and bottom panels). Note that while the initial value of $A$ for this subspace is 1, the contribution of other states sets the initial $A$ of the whole system to 1/2.}
    \label{fig:two-site-cartoon}
\end{figure}

Several aspects of the observed spin dynamics can be captured by a two-site model. Although states on two spin-1 sites span a 9-dimensional Hilbert space, we can reduce the spin dynamics to a beat note between two states. Since exchange interactions do not change the total magnetization $\sum_{i=1}^{N} S_{i}^{z}$, the Hilbert space factorizes to subspaces with the same total magnetization (although $S_i^z$ can differ within a subspace). Furthermore, the initial superposition state is symmetric between the  left and right wells, and any change in $A$ comes from the two coupled states: $|ab\rangle_L |ab\rangle_R$ and $(|aa\rangle_L |bb\rangle_R +|bb\rangle_L |aa\rangle_R)/\sqrt{2}$, whose values of $A$ are $2$ and $-1$ respectively
(Fig.~\ref{fig:two-site-cartoon}). By describing these two states as two poles on a Bloch sphere, we see that the initial state is represented by a vector pointing somewhere between the north pole and the equator with a vertical fictitious external field. The quench in $J$ and $D$ suddenly changes the strength and the orientation of this external field and induces a precession of the state vector around the new external field (see the Supplemental Material).  This results in an oscillation of $A$ with amplitude $2\left(D/J \right)/\left[ 9 + 4D/J + 4\left(D/J \right)^2 \right]$. This function has local extrema for $D/J = \pm 3/2$, but is not symmetric around $D/J = 0$. This explains the non-monotonic behaviour as a function of lattice depth, and shows why the contrast is smaller for positive $D/J$ than for negative $D/J$.

One would expect that for a larger number of sites, additional precession frequencies appear, turning the periodic oscillation for two sites into a relaxation toward an asymptotic value. Comparison between the two-site model and a many-site model numerically simulated using the time-evolution block-decimation algorithm for matrix-product states (MPS-TEBD) shows that the initial change in $A$ is indeed well captured by the two-site model (see the Supplemental Material). Due to the spin dynamics, the system evolves from a product state into a highly correlated state with entanglement between sites; this has been the focus of recent theoretical works \cite{Morera2019, Venegas-Gomez2020-2}. In the two-site model, the von Neumann entanglement entropy can reach up to $\sim0.9\times\ln\left(3\right)$ due to the interplay between single-ion anisotropy and exchange terms. This corresponds to an almost maximally entangled state since $\ln\left(3\right)$ is the maximum entropy for a spin-1 site.

\begin{figure}
    \centering
    \includegraphics[width=\columnwidth]{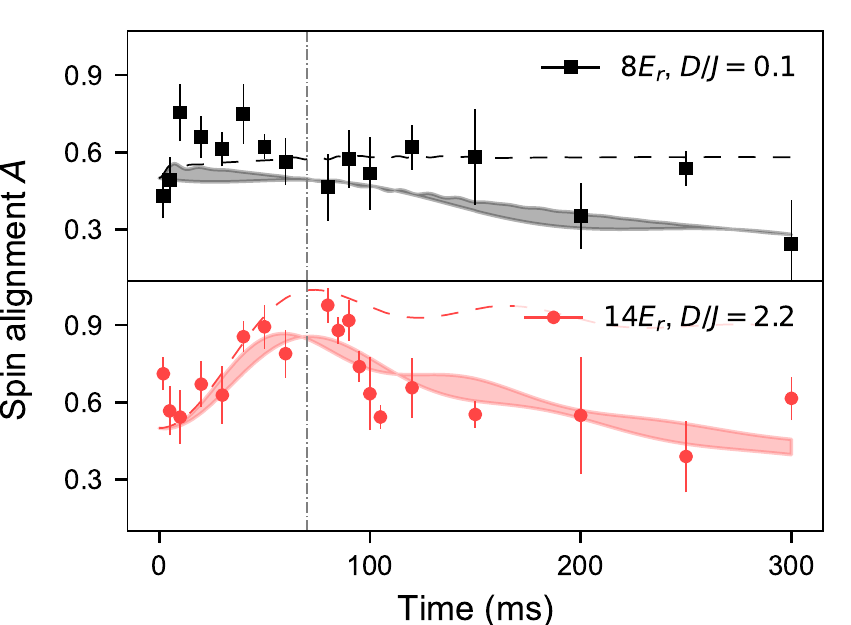}\\
    \includegraphics[width=\columnwidth]{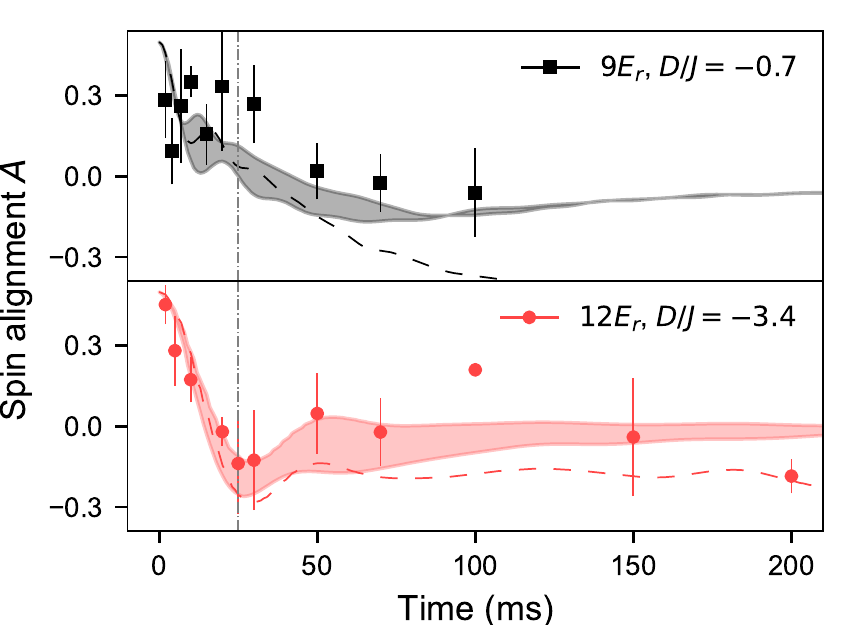}
    \caption{Coherent dynamics of the spin alignment $A$ after a quench in $D/J$. Varying the hold time at characteristic lattice depths for both positive and negative values of $D/J$ (top and bottom pairs of panels, respectively) reveals that strong transients in $A$ only occur at intermediate lattice depth for which $D$ and $J$ are comparable. The vertical, dash-dotted lines indicate the hold times used for these pairs in Fig.~\ref{fig:lattice-depth-scan}. Dashed lines are the results of the MPS-TEBD simulation. The shaded regions denote the MPS-TEBD results with $\pm 0.5~E_R$ uncertainty in the lattice depths, and include exponential decay towards a thermal spin state with $A=0$ with empirical $1/e$ times of $400~\mathrm{ms}$ ($D > 0$) and $100~\mathrm{ms}$ ($D < 0$).  The error bars are computed in the same manner as those in Fig.~\ref{fig:lattice-depth-scan}.}
    \label{fig:hold_time_scan}
\end{figure}

To illustrate that changes in the spin alignment $A$ result from competition between the exchange interaction and the single-ion anisotropy, we study the time evolution of $A$ at two different lattice depths (Fig.~\ref{fig:hold_time_scan}). For positive $D$, MPS-TEBD simulations predict very little change in $A$ at a lower lattice depth, where the exchange constant is relatively large, but the anisotropy is small, while it predicts a noticeable change in $A$ at a higher lattice depth, where the exchange constant and the anisotropy term becomes comparable.
While the simulation predicts equilibration of $A$ to an asymptotic value (thin lines), measurements show that it decays toward a lower value for positive $D$ and does not decrease as much as the simulation prediction for negative $D$. The measurements are consistent with the fact that at high spin temperature, the spin distribution becomes isotropic and $A$ vanishes. Indeed, when we ramp down the lattices and retrieve a Bose--Einstein condensate, we observe a significant reduction of condensate fraction after 300 ms.

In conclusion, we have implemented a spin-1 Heisenberg model with a single-ion anisotropy using the $\nu = 2$ plateau of a Mott insulator, and have observed the subtle interplay between spin exchange and on-site anisotropy in coherent spin dynamics. Much larger values of $D$ can be implemented with spin-dependent lattices, which will allow us to observe much faster anisotropy-driven dynamics, and will also enable mapping out the phase diagram of the anisotropic spin Hamiltonian \cite{Schachenmayer2015}.
It should also be noted that it is possible to change the sign of $J$ with the gradient of an optical dipole potential \cite{dimitrova2020,Sun2020}, which will permit exploration of the antiferromagnetic sector with bosons. Interesting dynamical features of anisotropic spin models have been predicted~\cite{Venegas-Gomez2020} including transient spin currents, implying counterflow superfluidity.

Regarding quantum simulations, single-ion anisotropies play a crucial role in magnetic materials (e.g.\ monolayers containing chromium \cite{gong2017discovery, Xu2018}). In such materials, crystal field effects lift the degeneracy of $d$-orbitals, and spin-orbit interaction transfers this anisotropy to the electronic spins responsible for the magnetism \cite{dai2008effects}. Here we have simulated this anisotropy by selecting a pair of atomic hyperfine states where the interspecies scattering length is different from the average of the intraspecies values.  This illustrates the potential for ultracold atoms in optical lattices to implement idealized Hamiltonians describing important materials.

\begin{acknowledgments}
We thank Colin Kennedy, William Cody Burton and Wenlan Chen for contributions to the development of experimental techniques, and Ivana Dimitrova for critical reading of the manuscript. We acknowledge support from the NSF through the Center for Ultracold Atoms and Grant No. 1506369, ARO-MURI Non-equilibrium Many-Body Dynamics (Grant No.\ W911NF14-1-0003), AFOSR-MURI Quantum Phases of Matter (Grant No. FA9550-14-1-0035), ONR (Grant No.\ N00014-17-1-2253), and a Vannevar-Bush Faculty Fellowship. W.C.C.\ acknowledges additional support from the Samsung Scholarship. 
\end{acknowledgments}

\appendix

\bibliography{spdf_oscillation.bib}

\input{supplemental_material}

\end{document}

%% file: supplemental_material.tex
\clearpage

\section*{Supplemental Material}

\subsection*{Calculation of $D$ and $J$}
The superexchange parameter $J$ and single-ion anisotropy $D$ were calculated using maximally localized Wannier functions for a simply cubic lattice \cite{Marzari1997, Kohn1959} and the scattering lengths in Table~\ref{tab:scattering-lengths}.

\begin{figure}
    \centering
    \includegraphics[width = \columnwidth]{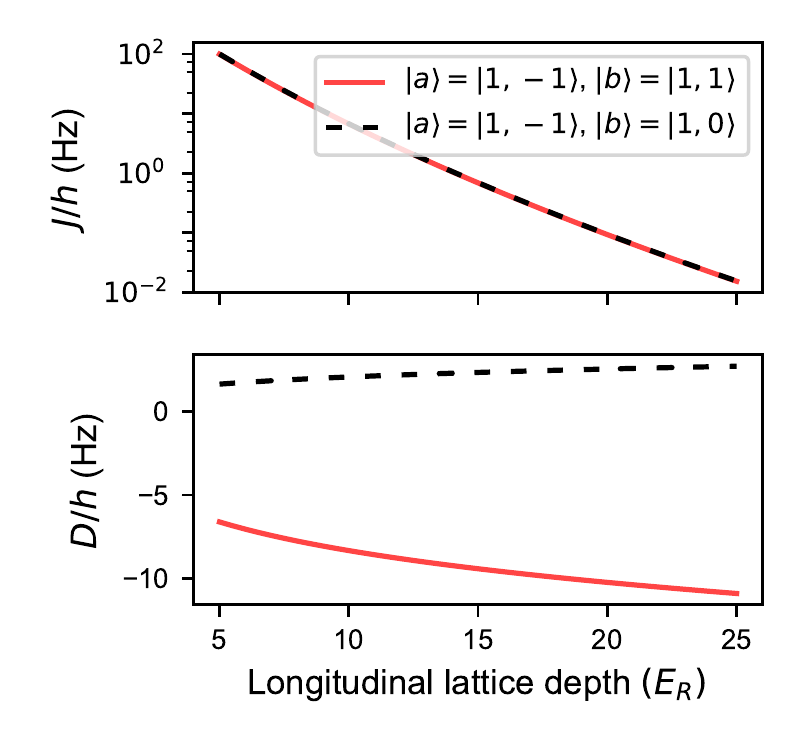}
    \caption{Values of $D$ and $J$ as a function of longitudinal lattice depth. The results are based on the scattering lengths given in Table~\ref{tab:scattering-lengths}, and assume transverse lattice depths of $35\,E_r$.}
    \label{fig:j-and-d-vs-lattice}
\end{figure}

The sign of $D$ is important for the qualitative behavior. Of the $F = 1$ states, the only combination with $D < 0$ is that of the $\ket{1, -1}$ and $\ket{1, 1}$ states. Any pair involving the $\ket{1, 0}$ state has a positive value of $D$; we chose the $\ket{1, -1}$ and $\ket{1, 0}$ combination because it was the easiest to prepare from the initial $\ket{1, -1}$ state. As mentioned in the main text, the value of $D$ is proportional to the various onsite interactions, which have a linear dependence on the scattering lengths. This means that $D \propto \left(a_{aa} + a_{bb}\right)/2 - a_{ab}$ which equals $-0.93 a_0$ and $0.23 a_0$ for the two chosen pairs. Through the Wannier functions, $D$ and $J$ depend on the lattice depth, which dependence is shown in Fig.~\ref{fig:j-and-d-vs-lattice}.

\subsection*{Confinement parameters}
The three-dimensional lattice is created by retro-reflecting three 1064-nm wavelength laser beams. The two horizontal beams have Gaussian beam waists of $150~\mathrm{\mu m}$, while the vertical lattice beam has a waist of $270~\mathrm{\mu m}$.  During the entire experiment the atoms are being held in a crossed-beam optical dipole trap. This consists of a vertical beam (which has isotropic trap frequencies of $2\pi \times 24~\mathrm{Hz}$) intersecting a highly elongated horizontal beam that is at a 45$^\circ$ angle with respect to the horizontal lattices. The latter primarily serves to hold the atoms against gravity, and it has trap frequencies of $2\pi \times 13~\mathrm{Hz}$ and $2\pi \times 130~\mathrm{Hz}$ along its horizontal and vertical axes, respectively. 

Using these parameters, we were able to calculate the occupation statistics of the Mott insulator, and obtained plateau fractions analogous to those presented in Ref.~\cite{Cheinet2008}. We desire a large $\nu = 2$ Mott insulator plateau, while avoiding any population in the $\nu = 3$ shell as that would interfere with the doublon measurements. Occasionally, we have monitored the population in the different shells using clock-shift spectroscopy \cite{Campbell2006}. On a day-to-day basis, however, we use the total atom number or the doublon fraction as indicators (note that our doublon detection scheme detects all the atoms on sites with $\nu \ge 2$.). To be safe, the doublon fraction is kept below $0.5$, and the atom number below $40\times 10^3$; for these parameters the population in $\nu = 3$ should be negligible.

\begin{table}
    \centering
    \begin{ruledtabular}
    \begin{tabular}{l|lll}
                     & $\ket{1,-1}$ & $\ket{1,0}$ & $\ket{1,1}$ \\ \hline
        $\ket{1,-1}$ & 100.4        & 100.4       & 101.333     \\
        $\ket{1,0}$  &              & 100.867     & 100.4       \\
        $\ket{1,1}$  &              &             & 100.4       \\
    \end{tabular}
    \end{ruledtabular}
    \caption{Scattering lengths in units of $a_0$ calculated using the values tabulated in Ref.~\cite{Stamper-Kurn13}.}
    \label{tab:scattering-lengths}
\end{table}

\subsection*{State preparation \& doublon measurement}
The initial state is prepared by a diabatic Landau--Zener sweep from the initial $\ket{1, -1}$ state to the $\ket{2, 0}$ state. The sweep parameters are set in such a way that we robustly create an equal superposition of the two states. Depending on whether we want to probe positive or negative $D/J$ we either transfer the population fraction in $\ket{2, 0}$ to $\ket{1, 0}$ using a $\pi$ pulse (which has small sensitivity to magnetic-field fluctuations), or to $\ket{1, 1}$ using an adiabatic Landau--Zener sweep.

As described in the main text, the doublon statistics are derived from three separate measurements of the atom number, two of them after inducing selective losses that depend on the doublon type. To measure the total doublon fraction, all doublons are removed, regardless of their internal states. Dipolar relaxation is too slow, so a Feshbach resonance between the $\ket{1, 1}$ and $\ket{2, 0}$ states can be used. For this, the $\ket{1, -1}$ component of the pair is transferred to the $\ket{2, 0}$ state using a Landau--Zener sweep, while the other pair component is left in or put into the $\ket{1, 1}$ state. The pairs are removed by modulating the magnetic bias field around the narrow Feshbach resonance at $9.045~\mathrm{G}$ \cite{kaufman2009radio}. Since the composition of the pairs we want to remove is arbitrary (they can be either $\ket{aa}$, $\ket{ab}$, or $\ket{bb}$), we employ a diabatic Landau--Zener sweep between $|2,0\rangle$ and $|1,1\rangle$ states simultaneously with the bias modulation, to make sure any doublon spends some time in the Feshbach pair state in order to be removed. In practice, a removal time of $80~\mathrm{ms}$ is sufficient.

In order to specifically remove \emph{paired} doublons (i.e.\ those of the $\ket{ab}$ type), we transfer the $\ket{1, -1}$ component of the pair to the $\ket{2, -2}$ state, and ensure that the other component is in the $\ket{1, 1}$ state. To remove these pairs, the bias field is modulated around the $9.092~\mathrm{G}$ Feshbach resonance between the $|1,1\rangle$ and $|2,-2\rangle$ states~\cite{kaufman2009radio}.

\subsection*{Two-site model}
In the limit of two sites, the spin Hamiltonian (\ref{eq:spin-1-Hamiltonian}) reduces to
\begin{equation}
    \mathcal{H} = -J \mathbf{S}_1\cdot\mathbf{S}_2 + D\left[ \left( S_1^z \right)^2 + \left( S_2^z \right)^2 \right].
\end{equation}
The initial state is a product state between site 1 and site 2: $\ket{\Psi} = \ket{\psi}_1\otimes\ket{\psi}_2$, where the single-site state is given by:
\begin{eqnarray}\label{eq:two-site-initial-state}
    \ket{\psi} &=& \left(\frac{|a\rangle - i |b\rangle}{\sqrt{2}}\right)_{\text{atom 1}}\otimes\left(\frac{|a\rangle - i |b\rangle}{\sqrt{2}}\right)_{\text{atom 2}} \nonumber\\
    &=& \frac{1}{2} \left( \ket{1} - i\sqrt{2} \ket{0} - \ket{-1} \right)
\end{eqnarray}
\comment[inline]{the signs of $\ket{1}$ and $\ket{-1}$ were flipped, to stay consistent with Fig.~\ref{fig:exp_procedure}. Check this doesn't affect the subsequent math.}
The full Hilbert space describing the two spin-1 sites is nine-dimensional.  However, the Hamiltonian is block diagonal in the total spin projection, $S_1^z + S_2^z$, and also with regard to odd and even symmetry between the two sites. For the state prepared initially, all the dynamics takes place in the symmetric $S_1^z + S_2^z = 0$ subspace, which contains only two states: $\left\{\left( \ket{1, -1} + \ket{-1, 1} \right)/\sqrt{2}, \ket{0, 0} \right\}$. The Hamiltonian is given by
\begin{equation}\label{eq:transformed-two-site-hamiltonian}
    \mathcal{H} = \begin{pmatrix}
    J + 2D & -\sqrt{2} J \\
    -\sqrt{2} J & 0
    \end{pmatrix}.
 \end{equation}

The projection of the initial state into this subspace is 
\begin{equation}\label{eq:two-site-reduced-wave-function}
    \ket{\psi} = \sqrt{\frac{1}{6}} \left( \ket{1, -1} + \ket{-1, 1} \right) + \sqrt{\frac{2}{3}} \ket{0, 0},
\end{equation}
also see Fig.~\ref{fig:two-site-cartoon}.  Since the $\ket{0,0}$ state has $\left(S^z_i\right)^2 = 0$, and the  $\left(\ket{1,-1} + \ket{-1, 1}\right)/\sqrt{2}$ has $\left(S^z_i\right)^2 = 1$, a Rabi oscillation between them leads to an oscillation of the spin alignment $A$.  Note that the components of the initial state in other subspaces contribute a constant value to $A$.

Inspection of the Hamiltonian (\ref{eq:transformed-two-site-hamiltonian}) identifies $J + 2D$ as a $z$ field, which is added to an $x$ field equal to $\sqrt{2}J$. In a deep lattice with $J \sim 0$, the field is parallel to the $z$ axis, but lowering the lattice adds an $x$ field, which tilts the field vector and initiates a precession of the state vector around it (see Fig.~\ref{fig:two-site-cartoon}).

\begin{figure}
    \centering
    \includegraphics[width=\columnwidth]{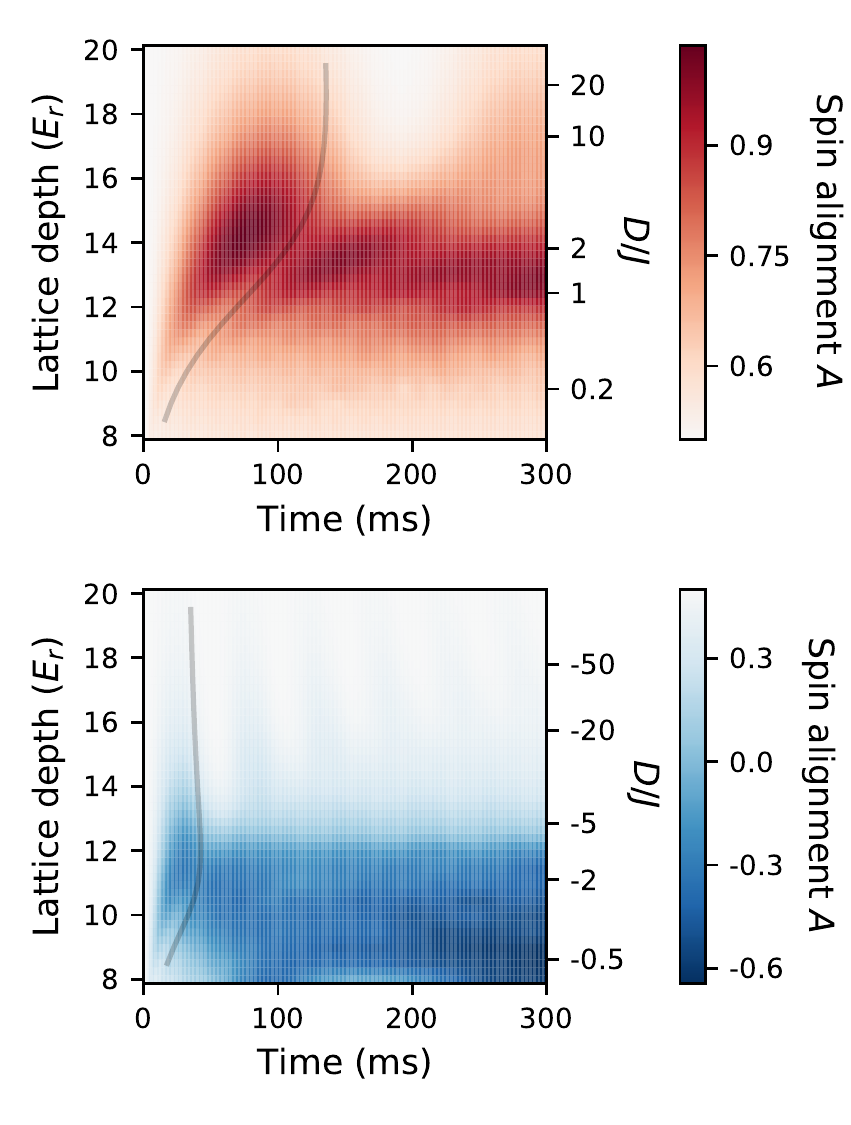}
    \caption{Time evolution of the spin alignment $A$ for various lattice depths, calculated using the TEBD algorithm for matrix-product states. The top and bottom figures are calculated for pairs with positive ($\ket{1,-1}$ and $\ket{1, 0}$) and negative ($\ket{1,-1}$ and $\ket{1,1}$) values of $D/J$, respectively. The solid lines indicate the inverse (lattice depth dependent) Rabi frequency of Eq.~(\ref{eq:two-site-rabi}) times $\pi\sqrt{2}$, showing that the initial behavior is dictated by nearest neighbors.}
    \label{fig:full-mps-results}
\end{figure}

The Rabi frequency of this oscillation is given by
\begin{equation}\label{eq:two-site-rabi}
    \Omega = \sqrt{9J^2 + 4JD + 4D^2}/\hbar ,
\end{equation}
while the amplitude of the oscillation in $A$ is $2JD/\Omega^2$, which is maximized for $|D/J| = 3/2$.

\subsection*{Matrix-product state simulations}
We implemented the time-evolving block decimation algorithm for matrix-product states (MPS-TEBD) \cite{Hauschild2018, Vidal2004} on 100 sites, using a maximum bond dimension of 20. This was found to give results consistent with published data  \cite{Venegas-Gomez2020}. The modest bond dimension is sufficient because the transient behavior in $A$ occurs within a few exchange times ($\hbar/J$), during which correlations only build up between clusters of sites. This has the additional benefit that the calculation can be run on a desktop computer.

The simulated evolution of the spin alignment $A$ as a function of lattice depth is shown in Fig.~\ref{fig:full-mps-results}. These results form the basis of the simulations presented in the main text. Comparing to the two-site model, we observe that the early time behavior is dominated by nearest-neighbor physics. Specifically, the first minima seen in Fig.~\ref{fig:full-mps-results} occur at the period of the Rabi oscillation given in Eq.~(\ref{eq:two-site-rabi}) divided by $\sqrt{2}$ to account for the fact that a site in the chain has not one but two neighbors.

This also allows us to understand the choice of hold times in Fig.~\ref{fig:lattice-depth-scan} as $\pi/\sqrt{2}\Omega$ at the lattice depth where $|D/J| \sim 3/2$. This number equals $67$ and $17~\mathrm{ms}$ for the positive and negative $D/J$ pair, respectively, while the actual values we use are $70$ and $25~\mathrm{ms}$. It can be thought of as half a Rabi oscillation to ensure the largest possible contrast in the signal.

%% file: spdf_oscillation.bbl
\begin{thebibliography}{35}%
\makeatletter
\providecommand \@ifxundefined [1]{%
 \@ifx{#1\undefined}
}%
\providecommand \@ifnum [1]{%
 \ifnum #1\expandafter \@firstoftwo
 \else \expandafter \@secondoftwo
 \fi
}%
\providecommand \@ifx [1]{%
 \ifx #1\expandafter \@firstoftwo
 \else \expandafter \@secondoftwo
 \fi
}%
\providecommand \natexlab [1]{#1}%
\providecommand \enquote  [1]{``#1''}%
\providecommand \bibnamefont  [1]{#1}%
\providecommand \bibfnamefont [1]{#1}%
\providecommand \citenamefont [1]{#1}%
\providecommand \href@noop [0]{\@secondoftwo}%
\providecommand \href [0]{\begingroup \@sanitize@url \@href}%
\providecommand \@href[1]{\@@startlink{#1}\@@href}%
\providecommand \@@href[1]{\endgroup#1\@@endlink}%
\providecommand \@sanitize@url [0]{\catcode `\\12\catcode `\$12\catcode
  `\&12\catcode `\#12\catcode `\^12\catcode `\_12\catcode `\%12\relax}%
\providecommand \@@startlink[1]{}%
\providecommand \@@endlink[0]{}%
\providecommand \url  [0]{\begingroup\@sanitize@url \@url }%
\providecommand \@url [1]{\endgroup\@href {#1}{\urlprefix }}%
\providecommand \urlprefix  [0]{URL }%
\providecommand \Eprint [0]{\href }%
\providecommand \doibase [0]{http://dx.doi.org/}%
\providecommand \selectlanguage [0]{\@gobble}%
\providecommand \bibinfo  [0]{\@secondoftwo}%
\providecommand \bibfield  [0]{\@secondoftwo}%
\providecommand \translation [1]{[#1]}%
\providecommand \BibitemOpen [0]{}%
\providecommand \bibitemStop [0]{}%
\providecommand \bibitemNoStop [0]{.\EOS\space}%
\providecommand \EOS [0]{\spacefactor3000\relax}%
\providecommand \BibitemShut  [1]{\csname bibitem#1\endcsname}%
\let\auto@bib@innerbib\@empty
\bibitem [{\citenamefont {Bloch}\ \emph {et~al.}(2008)\citenamefont {Bloch},
  \citenamefont {Dalibard},\ and\ \citenamefont {Zwerger}}]{bloch2008many}%
  \BibitemOpen
  \bibfield  {author} {\bibinfo {author} {\bibfnamefont {I.}~\bibnamefont
  {Bloch}}, \bibinfo {author} {\bibfnamefont {J.}~\bibnamefont {Dalibard}}, \
  and\ \bibinfo {author} {\bibfnamefont {W.}~\bibnamefont {Zwerger}},\
  }\href@noop {} {\bibfield  {journal} {\bibinfo  {journal} {Reviews of modern
  physics}\ }\textbf {\bibinfo {volume} {80}},\ \bibinfo {pages} {885}
  (\bibinfo {year} {2008})}\BibitemShut {NoStop}%
\bibitem [{\citenamefont {Duan}\ \emph {et~al.}(2003)\citenamefont {Duan},
  \citenamefont {Demler},\ and\ \citenamefont {Lukin}}]{duan2003controlling}%
  \BibitemOpen
  \bibfield  {author} {\bibinfo {author} {\bibfnamefont {L.-M.}\ \bibnamefont
  {Duan}}, \bibinfo {author} {\bibfnamefont {E.}~\bibnamefont {Demler}}, \ and\
  \bibinfo {author} {\bibfnamefont {M.~D.}\ \bibnamefont {Lukin}},\ }\href@noop
  {} {\bibfield  {journal} {\bibinfo  {journal} {Physical Review Letters}\
  }\textbf {\bibinfo {volume} {91}},\ \bibinfo {pages} {090402} (\bibinfo
  {year} {2003})}\BibitemShut {NoStop}%
\bibitem [{\citenamefont {Kuklov}\ and\ \citenamefont
  {Svistunov}(2003)}]{kuklov2003counterflow}%
  \BibitemOpen
  \bibfield  {author} {\bibinfo {author} {\bibfnamefont {A.~B.}\ \bibnamefont
  {Kuklov}}\ and\ \bibinfo {author} {\bibfnamefont {B.~V.}\ \bibnamefont
  {Svistunov}},\ }\href {\doibase 10.1103/PhysRevLett.90.100401} {\bibfield
  {journal} {\bibinfo  {journal} {Phys. Rev. Lett.}\ }\textbf {\bibinfo
  {volume} {90}},\ \bibinfo {pages} {100401} (\bibinfo {year}
  {2003})}\BibitemShut {NoStop}%
\bibitem [{\citenamefont {Mazurenko}\ \emph {et~al.}(2017)\citenamefont
  {Mazurenko}, \citenamefont {Chiu}, \citenamefont {Ji}, \citenamefont
  {Parsons}, \citenamefont {Kan\'asz-Nagy}, \citenamefont {Schmidt},
  \citenamefont {Grusdt}, \citenamefont {Demler}, \citenamefont {Greiff},\ and\
  \citenamefont {Greiner}}]{Mazurenko2017}%
  \BibitemOpen
  \bibfield  {author} {\bibinfo {author} {\bibfnamefont {A.}~\bibnamefont
  {Mazurenko}}, \bibinfo {author} {\bibfnamefont {C.~S.}\ \bibnamefont {Chiu}},
  \bibinfo {author} {\bibfnamefont {G.}~\bibnamefont {Ji}}, \bibinfo {author}
  {\bibfnamefont {M.~F.}\ \bibnamefont {Parsons}}, \bibinfo {author}
  {\bibfnamefont {M.}~\bibnamefont {Kan\'asz-Nagy}}, \bibinfo {author}
  {\bibfnamefont {R.}~\bibnamefont {Schmidt}}, \bibinfo {author} {\bibfnamefont
  {F.}~\bibnamefont {Grusdt}}, \bibinfo {author} {\bibfnamefont
  {E.}~\bibnamefont {Demler}}, \bibinfo {author} {\bibfnamefont
  {D.}~\bibnamefont {Greiff}}, \ and\ \bibinfo {author} {\bibfnamefont
  {M.}~\bibnamefont {Greiner}},\ }\href@noop {} {\bibfield  {journal} {\bibinfo
   {journal} {Nature}\ }\textbf {\bibinfo {volume} {545}},\ \bibinfo {pages}
  {462} (\bibinfo {year} {2017})}\BibitemShut {NoStop}%
\bibitem [{\citenamefont {Jepsen}\ \emph {et~al.}(2020)\citenamefont {Jepsen},
  \citenamefont {Amato-Grill}, \citenamefont {Dimitrova}, \citenamefont {Ho},
  \citenamefont {Demler},\ and\ \citenamefont {Ketterle}}]{jepsen2020}%
  \BibitemOpen
  \bibfield  {author} {\bibinfo {author} {\bibfnamefont {P.~N.}\ \bibnamefont
  {Jepsen}}, \bibinfo {author} {\bibfnamefont {J.}~\bibnamefont {Amato-Grill}},
  \bibinfo {author} {\bibfnamefont {I.}~\bibnamefont {Dimitrova}}, \bibinfo
  {author} {\bibfnamefont {W.~W.}\ \bibnamefont {Ho}}, \bibinfo {author}
  {\bibfnamefont {E.}~\bibnamefont {Demler}}, \ and\ \bibinfo {author}
  {\bibfnamefont {W.}~\bibnamefont {Ketterle}},\ }\href@noop {} {\bibfield
  {journal} {\bibinfo  {journal} {Nature}\ }\textbf {\bibinfo {volume} {588}},\
  \bibinfo {pages} {403} (\bibinfo {year} {2020})}\BibitemShut {NoStop}%
\bibitem [{\citenamefont {de~Paz}\ \emph {et~al.}(2013)\citenamefont {de~Paz},
  \citenamefont {Sharma}, \citenamefont {Chotia}, \citenamefont {Mar\'echal},
  \citenamefont {Huckans}, \citenamefont {Pedri}, \citenamefont {Santos},
  \citenamefont {Gorceix}, \citenamefont {Vernac},\ and\ \citenamefont
  {Laburthe-Tolra}}]{dePaz2013}%
  \BibitemOpen
  \bibfield  {author} {\bibinfo {author} {\bibfnamefont {A.}~\bibnamefont
  {de~Paz}}, \bibinfo {author} {\bibfnamefont {A.}~\bibnamefont {Sharma}},
  \bibinfo {author} {\bibfnamefont {A.}~\bibnamefont {Chotia}}, \bibinfo
  {author} {\bibfnamefont {E.}~\bibnamefont {Mar\'echal}}, \bibinfo {author}
  {\bibfnamefont {J.~H.}\ \bibnamefont {Huckans}}, \bibinfo {author}
  {\bibfnamefont {P.}~\bibnamefont {Pedri}}, \bibinfo {author} {\bibfnamefont
  {L.}~\bibnamefont {Santos}}, \bibinfo {author} {\bibfnamefont
  {O.}~\bibnamefont {Gorceix}}, \bibinfo {author} {\bibfnamefont
  {L.}~\bibnamefont {Vernac}}, \ and\ \bibinfo {author} {\bibfnamefont
  {B.}~\bibnamefont {Laburthe-Tolra}},\ }\href {\doibase
  10.1103/PhysRevLett.111.185305} {\bibfield  {journal} {\bibinfo  {journal}
  {Phys. Rev. Lett.}\ }\textbf {\bibinfo {volume} {111}},\ \bibinfo {pages}
  {185305} (\bibinfo {year} {2013})}\BibitemShut {NoStop}%
\bibitem [{\citenamefont {Kitagawa}\ and\ \citenamefont
  {Ueda}(1993)}]{KitagawaUeda}%
  \BibitemOpen
  \bibfield  {author} {\bibinfo {author} {\bibfnamefont {M.}~\bibnamefont
  {Kitagawa}}\ and\ \bibinfo {author} {\bibfnamefont {M.}~\bibnamefont
  {Ueda}},\ }\href {\doibase 10.1103/PhysRevA.47.5138} {\bibfield  {journal}
  {\bibinfo  {journal} {Phys. Rev. A}\ }\textbf {\bibinfo {volume} {47}},\
  \bibinfo {pages} {5138} (\bibinfo {year} {1993})}\BibitemShut {NoStop}%
\bibitem [{\citenamefont {Li}\ \emph {et~al.}(2011)\citenamefont {Li},
  \citenamefont {Bakhtiari}, \citenamefont {He},\ and\ \citenamefont
  {Hofstetter}}]{Li2011}%
  \BibitemOpen
  \bibfield  {author} {\bibinfo {author} {\bibfnamefont {Y.}~\bibnamefont
  {Li}}, \bibinfo {author} {\bibfnamefont {M.~R.}\ \bibnamefont {Bakhtiari}},
  \bibinfo {author} {\bibfnamefont {L.}~\bibnamefont {He}}, \ and\ \bibinfo
  {author} {\bibfnamefont {W.}~\bibnamefont {Hofstetter}},\ }\href {\doibase
  10.1103/PhysRevB.84.144411} {\bibfield  {journal} {\bibinfo  {journal} {Phys.
  Rev. B}\ }\textbf {\bibinfo {volume} {84}},\ \bibinfo {pages} {144411}
  (\bibinfo {year} {2011})}\BibitemShut {NoStop}%
\bibitem [{\citenamefont {Li}\ \emph {et~al.}(2016)\citenamefont {Li},
  \citenamefont {He},\ and\ \citenamefont {Hofstetter}}]{Li2016}%
  \BibitemOpen
  \bibfield  {author} {\bibinfo {author} {\bibfnamefont {Y.}~\bibnamefont
  {Li}}, \bibinfo {author} {\bibfnamefont {L.}~\bibnamefont {He}}, \ and\
  \bibinfo {author} {\bibfnamefont {W.}~\bibnamefont {Hofstetter}},\ }\href
  {\doibase 10.1103/PhysRevA.93.033622} {\bibfield  {journal} {\bibinfo
  {journal} {Phys. Rev. A}\ }\textbf {\bibinfo {volume} {93}},\ \bibinfo
  {pages} {033622} (\bibinfo {year} {2016})}\BibitemShut {NoStop}%
\bibitem [{\citenamefont {Altman}\ \emph {et~al.}(2003)\citenamefont {Altman},
  \citenamefont {Hofstetter}, \citenamefont {Demler},\ and\ \citenamefont
  {Lukin}}]{altman2003phase}%
  \BibitemOpen
  \bibfield  {author} {\bibinfo {author} {\bibfnamefont {E.}~\bibnamefont
  {Altman}}, \bibinfo {author} {\bibfnamefont {W.}~\bibnamefont {Hofstetter}},
  \bibinfo {author} {\bibfnamefont {E.}~\bibnamefont {Demler}}, \ and\ \bibinfo
  {author} {\bibfnamefont {M.~D.}\ \bibnamefont {Lukin}},\ }\href@noop {}
  {\bibfield  {journal} {\bibinfo  {journal} {New Journal of Physics}\ }\textbf
  {\bibinfo {volume} {5}},\ \bibinfo {pages} {113} (\bibinfo {year}
  {2003})}\BibitemShut {NoStop}%
\bibitem [{\citenamefont {Schachenmayer}\ \emph {et~al.}(2015)\citenamefont
  {Schachenmayer}, \citenamefont {Weld}, \citenamefont {Miyake}, \citenamefont
  {Siviloglou}, \citenamefont {Ketterle},\ and\ \citenamefont
  {Daley}}]{Schachenmayer2015}%
  \BibitemOpen
  \bibfield  {author} {\bibinfo {author} {\bibfnamefont {J.}~\bibnamefont
  {Schachenmayer}}, \bibinfo {author} {\bibfnamefont {D.~M.}\ \bibnamefont
  {Weld}}, \bibinfo {author} {\bibfnamefont {H.}~\bibnamefont {Miyake}},
  \bibinfo {author} {\bibfnamefont {G.~A.}\ \bibnamefont {Siviloglou}},
  \bibinfo {author} {\bibfnamefont {W.}~\bibnamefont {Ketterle}}, \ and\
  \bibinfo {author} {\bibfnamefont {A.~J.}\ \bibnamefont {Daley}},\ }\href
  {\doibase 10.1103/PhysRevA.92.041602} {\bibfield  {journal} {\bibinfo
  {journal} {Phys. Rev. A}\ }\textbf {\bibinfo {volume} {92}},\ \bibinfo
  {pages} {041602(R)} (\bibinfo {year} {2015})}\BibitemShut {NoStop}%
\bibitem [{\citenamefont {Haldane}(1983{\natexlab{a}})}]{Haldane1983}%
  \BibitemOpen
  \bibfield  {author} {\bibinfo {author} {\bibfnamefont {F.~D.~M.}\
  \bibnamefont {Haldane}},\ }\href {\doibase
  https://doi.org/10.1016/0375-9601(83)90631-X} {\bibfield  {journal} {\bibinfo
   {journal} {Physics Letters A}\ }\textbf {\bibinfo {volume} {93}},\ \bibinfo
  {pages} {464 } (\bibinfo {year} {1983}{\natexlab{a}})}\BibitemShut {NoStop}%
\bibitem [{\citenamefont {Haldane}(1983{\natexlab{b}})}]{Haldane1983PRL}%
  \BibitemOpen
  \bibfield  {author} {\bibinfo {author} {\bibfnamefont {F.~D.~M.}\
  \bibnamefont {Haldane}},\ }\href {\doibase 10.1103/PhysRevLett.50.1153}
  {\bibfield  {journal} {\bibinfo  {journal} {Phys. Rev. Lett.}\ }\textbf
  {\bibinfo {volume} {50}},\ \bibinfo {pages} {1153} (\bibinfo {year}
  {1983}{\natexlab{b}})}\BibitemShut {NoStop}%
\bibitem [{\citenamefont {Haldane}(2017)}]{Haldane2017}%
  \BibitemOpen
  \bibfield  {author} {\bibinfo {author} {\bibfnamefont {F.~D.~M.}\
  \bibnamefont {Haldane}},\ }\href {\doibase 10.1103/RevModPhys.89.040502}
  {\bibfield  {journal} {\bibinfo  {journal} {Rev. Mod. Phys.}\ }\textbf
  {\bibinfo {volume} {89}},\ \bibinfo {pages} {040502} (\bibinfo {year}
  {2017})}\BibitemShut {NoStop}%
\bibitem [{\citenamefont {Mermin}\ and\ \citenamefont
  {Wagner}(1966)}]{merminwagner1966}%
  \BibitemOpen
  \bibfield  {author} {\bibinfo {author} {\bibfnamefont {N.~D.}\ \bibnamefont
  {Mermin}}\ and\ \bibinfo {author} {\bibfnamefont {H.}~\bibnamefont
  {Wagner}},\ }\href {\doibase 10.1103/PhysRevLett.17.1133} {\bibfield
  {journal} {\bibinfo  {journal} {Phys. Rev. Lett.}\ }\textbf {\bibinfo
  {volume} {17}},\ \bibinfo {pages} {1133} (\bibinfo {year}
  {1966})}\BibitemShut {NoStop}%
\bibitem [{\citenamefont {Stre\v{c}ka}\ \emph {et~al.}(2008)\citenamefont
  {Stre\v{c}ka}, \citenamefont {J\'an},\ and\ \citenamefont
  {\v{C}anov\'a}}]{Strecka2008}%
  \BibitemOpen
  \bibfield  {author} {\bibinfo {author} {\bibfnamefont {J.}~\bibnamefont
  {Stre\v{c}ka}}, \bibinfo {author} {\bibfnamefont {D.}~\bibnamefont {J\'an}},
  \ and\ \bibinfo {author} {\bibfnamefont {L.}~\bibnamefont {\v{C}anov\'a}},\
  }\href@noop {} {\bibfield  {journal} {\bibinfo  {journal} {Chinese Journal of
  Physics}\ }\textbf {\bibinfo {volume} {46}},\ \bibinfo {pages} {329}
  (\bibinfo {year} {2008})}\BibitemShut {NoStop}%
\bibitem [{\citenamefont {Renard}\ \emph {et~al.}(1987)\citenamefont {Renard},
  \citenamefont {Verdaguer}, \citenamefont {Regnault}, \citenamefont
  {Erkelens}, \citenamefont {Rossat-Mignod},\ and\ \citenamefont
  {Stirling}}]{Renard1987}%
  \BibitemOpen
  \bibfield  {author} {\bibinfo {author} {\bibfnamefont {J.~P.}\ \bibnamefont
  {Renard}}, \bibinfo {author} {\bibfnamefont {M.}~\bibnamefont {Verdaguer}},
  \bibinfo {author} {\bibfnamefont {L.~P.}\ \bibnamefont {Regnault}}, \bibinfo
  {author} {\bibfnamefont {W.~A.~C.}\ \bibnamefont {Erkelens}}, \bibinfo
  {author} {\bibfnamefont {J.}~\bibnamefont {Rossat-Mignod}}, \ and\ \bibinfo
  {author} {\bibfnamefont {W.~G.}\ \bibnamefont {Stirling}},\ }\href {\doibase
  10.1209/0295-5075/3/8/013} {\bibfield  {journal} {\bibinfo  {journal}
  {Europhysics Letters ({EPL})}\ }\textbf {\bibinfo {volume} {3}},\ \bibinfo
  {pages} {945} (\bibinfo {year} {1987})}\BibitemShut {NoStop}%
\bibitem [{\citenamefont {Chauhan}\ \emph {et~al.}(2020)\citenamefont
  {Chauhan}, \citenamefont {Mahmood}, \citenamefont {Changlani}, \citenamefont
  {Koohpayeh},\ and\ \citenamefont {Armitage}}]{Chauhan2020}%
  \BibitemOpen
  \bibfield  {author} {\bibinfo {author} {\bibfnamefont {P.}~\bibnamefont
  {Chauhan}}, \bibinfo {author} {\bibfnamefont {F.}~\bibnamefont {Mahmood}},
  \bibinfo {author} {\bibfnamefont {H.~J.}\ \bibnamefont {Changlani}}, \bibinfo
  {author} {\bibfnamefont {S.~M.}\ \bibnamefont {Koohpayeh}}, \ and\ \bibinfo
  {author} {\bibfnamefont {N.~P.}\ \bibnamefont {Armitage}},\ }\href {\doibase
  10.1103/PhysRevLett.124.037203} {\bibfield  {journal} {\bibinfo  {journal}
  {Phys. Rev. Lett.}\ }\textbf {\bibinfo {volume} {124}},\ \bibinfo {pages}
  {037203} (\bibinfo {year} {2020})}\BibitemShut {NoStop}%
\bibitem [{\citenamefont {Senko}\ \emph {et~al.}(2015)\citenamefont {Senko},
  \citenamefont {Richerme}, \citenamefont {Smith}, \citenamefont {Lee},
  \citenamefont {Cohen}, \citenamefont {Retzker},\ and\ \citenamefont
  {Monroe}}]{Senko2015}%
  \BibitemOpen
  \bibfield  {author} {\bibinfo {author} {\bibfnamefont {C.}~\bibnamefont
  {Senko}}, \bibinfo {author} {\bibfnamefont {P.}~\bibnamefont {Richerme}},
  \bibinfo {author} {\bibfnamefont {J.}~\bibnamefont {Smith}}, \bibinfo
  {author} {\bibfnamefont {A.}~\bibnamefont {Lee}}, \bibinfo {author}
  {\bibfnamefont {I.}~\bibnamefont {Cohen}}, \bibinfo {author} {\bibfnamefont
  {A.}~\bibnamefont {Retzker}}, \ and\ \bibinfo {author} {\bibfnamefont
  {C.}~\bibnamefont {Monroe}},\ }\href {\doibase 10.1103/PhysRevX.5.021026}
  {\bibfield  {journal} {\bibinfo  {journal} {Phys. Rev. X}\ }\textbf {\bibinfo
  {volume} {5}},\ \bibinfo {pages} {021026} (\bibinfo {year}
  {2015})}\BibitemShut {NoStop}%
\bibitem [{\citenamefont {Kaufman}\ \emph {et~al.}(2009)\citenamefont
  {Kaufman}, \citenamefont {Anderson}, \citenamefont {Hanna}, \citenamefont
  {Tiesinga}, \citenamefont {Julienne},\ and\ \citenamefont
  {Hall}}]{kaufman2009radio}%
  \BibitemOpen
  \bibfield  {author} {\bibinfo {author} {\bibfnamefont {A.~M.}\ \bibnamefont
  {Kaufman}}, \bibinfo {author} {\bibfnamefont {R.~P.}\ \bibnamefont
  {Anderson}}, \bibinfo {author} {\bibfnamefont {T.~M.}\ \bibnamefont {Hanna}},
  \bibinfo {author} {\bibfnamefont {E.}~\bibnamefont {Tiesinga}}, \bibinfo
  {author} {\bibfnamefont {P.~S.}\ \bibnamefont {Julienne}}, \ and\ \bibinfo
  {author} {\bibfnamefont {D.~S.}\ \bibnamefont {Hall}},\ }\href {\doibase
  10.1103/PhysRevA.80.050701} {\bibfield  {journal} {\bibinfo  {journal} {Phys.
  Rev. A}\ }\textbf {\bibinfo {volume} {80}},\ \bibinfo {pages} {050701(R)}
  (\bibinfo {year} {2009})}\BibitemShut {NoStop}%
\bibitem [{\citenamefont {Morera}\ \emph {et~al.}(2019)\citenamefont {Morera},
  \citenamefont {Polls},\ and\ \citenamefont {Juliá-Díaz}}]{Morera2019}%
  \BibitemOpen
  \bibfield  {author} {\bibinfo {author} {\bibfnamefont {I.}~\bibnamefont
  {Morera}}, \bibinfo {author} {\bibfnamefont {A.}~\bibnamefont {Polls}}, \
  and\ \bibinfo {author} {\bibfnamefont {B.}~\bibnamefont {Juliá-Díaz}},\
  }\href@noop {} {\bibfield  {journal} {\bibinfo  {journal} {Scientific
  Reports}\ }\textbf {\bibinfo {volume} {9}},\ \bibinfo {pages} {9424}
  (\bibinfo {year} {2019})}\BibitemShut {NoStop}%
\bibitem [{\citenamefont {Venegas-Gomez}\ \emph
  {et~al.}(2020{\natexlab{a}})\citenamefont {Venegas-Gomez}, \citenamefont
  {Schachenmayer}, \citenamefont {Buyskikh}, \citenamefont {Ketterle},
  \citenamefont {Chiofalo},\ and\ \citenamefont {Daley}}]{Venegas-Gomez2020-2}%
  \BibitemOpen
  \bibfield  {author} {\bibinfo {author} {\bibfnamefont {A.}~\bibnamefont
  {Venegas-Gomez}}, \bibinfo {author} {\bibfnamefont {J.}~\bibnamefont
  {Schachenmayer}}, \bibinfo {author} {\bibfnamefont {A.~S.}\ \bibnamefont
  {Buyskikh}}, \bibinfo {author} {\bibfnamefont {W.}~\bibnamefont {Ketterle}},
  \bibinfo {author} {\bibfnamefont {M.~L.}\ \bibnamefont {Chiofalo}}, \ and\
  \bibinfo {author} {\bibfnamefont {A.~J.}\ \bibnamefont {Daley}},\ }\href@noop
  {} {\bibfield  {journal} {\bibinfo  {journal} {Quantum Science and
  Technology}\ }\textbf {\bibinfo {volume} {5}},\ \bibinfo {pages} {045013}
  (\bibinfo {year} {2020}{\natexlab{a}})}\BibitemShut {NoStop}%
\bibitem [{\citenamefont {Dimitrova}\ \emph {et~al.}(2020)\citenamefont
  {Dimitrova}, \citenamefont {Jepsen}, \citenamefont {Buyskikh}, \citenamefont
  {Venegas-Gomez}, \citenamefont {Amato-Grill}, \citenamefont {Daley},\ and\
  \citenamefont {Ketterle}}]{dimitrova2020}%
  \BibitemOpen
  \bibfield  {author} {\bibinfo {author} {\bibfnamefont {I.}~\bibnamefont
  {Dimitrova}}, \bibinfo {author} {\bibfnamefont {N.}~\bibnamefont {Jepsen}},
  \bibinfo {author} {\bibfnamefont {A.}~\bibnamefont {Buyskikh}}, \bibinfo
  {author} {\bibfnamefont {A.}~\bibnamefont {Venegas-Gomez}}, \bibinfo {author}
  {\bibfnamefont {J.}~\bibnamefont {Amato-Grill}}, \bibinfo {author}
  {\bibfnamefont {A.}~\bibnamefont {Daley}}, \ and\ \bibinfo {author}
  {\bibfnamefont {W.}~\bibnamefont {Ketterle}},\ }\href {\doibase
  10.1103/PhysRevLett.124.043204} {\bibfield  {journal} {\bibinfo  {journal}
  {Phys. Rev. Lett.}\ }\textbf {\bibinfo {volume} {124}},\ \bibinfo {pages}
  {043204} (\bibinfo {year} {2020})}\BibitemShut {NoStop}%
\bibitem [{\citenamefont {Sun}\ \emph {et~al.}(2020)\citenamefont {Sun},
  \citenamefont {Yang}, \citenamefont {Wang}, \citenamefont {Zhou},
  \citenamefont {Su}, \citenamefont {Dai}, \citenamefont {Yuan},\ and\
  \citenamefont {Pan}}]{Sun2020}%
  \BibitemOpen
  \bibfield  {author} {\bibinfo {author} {\bibfnamefont {H.}~\bibnamefont
  {Sun}}, \bibinfo {author} {\bibfnamefont {B.}~\bibnamefont {Yang}}, \bibinfo
  {author} {\bibfnamefont {H.-Y.}\ \bibnamefont {Wang}}, \bibinfo {author}
  {\bibfnamefont {Z.-Y.}\ \bibnamefont {Zhou}}, \bibinfo {author}
  {\bibfnamefont {G.-X.}\ \bibnamefont {Su}}, \bibinfo {author} {\bibfnamefont
  {H.-N.}\ \bibnamefont {Dai}}, \bibinfo {author} {\bibfnamefont {Z.-S.}\
  \bibnamefont {Yuan}}, \ and\ \bibinfo {author} {\bibfnamefont {J.-W.}\
  \bibnamefont {Pan}},\ }\href@noop {} {\bibfield  {journal} {\bibinfo
  {journal} {arXiv:2009.01426}\ } (\bibinfo {year} {2020})}\BibitemShut
  {NoStop}%
\bibitem [{\citenamefont {Venegas-Gomez}\ \emph
  {et~al.}(2020{\natexlab{b}})\citenamefont {Venegas-Gomez}, \citenamefont
  {Buyskikh}, \citenamefont {Schachenmayer}, \citenamefont {Ketterle},\ and\
  \citenamefont {Daley}}]{Venegas-Gomez2020}%
  \BibitemOpen
  \bibfield  {author} {\bibinfo {author} {\bibfnamefont {A.}~\bibnamefont
  {Venegas-Gomez}}, \bibinfo {author} {\bibfnamefont {A.~S.}\ \bibnamefont
  {Buyskikh}}, \bibinfo {author} {\bibfnamefont {J.}~\bibnamefont
  {Schachenmayer}}, \bibinfo {author} {\bibfnamefont {W.}~\bibnamefont
  {Ketterle}}, \ and\ \bibinfo {author} {\bibfnamefont {A.~J.}\ \bibnamefont
  {Daley}},\ }\href {\doibase 10.1103/PhysRevA.102.023321} {\bibfield
  {journal} {\bibinfo  {journal} {Phys. Rev. A}\ }\textbf {\bibinfo {volume}
  {102}},\ \bibinfo {pages} {023321} (\bibinfo {year}
  {2020}{\natexlab{b}})}\BibitemShut {NoStop}%
\bibitem [{\citenamefont {Gong}\ \emph {et~al.}(2017)\citenamefont {Gong},
  \citenamefont {Li}, \citenamefont {Li}, \citenamefont {Ji}, \citenamefont
  {Stern}, \citenamefont {Xia}, \citenamefont {Cao}, \citenamefont {Bao},
  \citenamefont {Wang}, \citenamefont {Wang} \emph
  {et~al.}}]{gong2017discovery}%
  \BibitemOpen
  \bibfield  {author} {\bibinfo {author} {\bibfnamefont {C.}~\bibnamefont
  {Gong}}, \bibinfo {author} {\bibfnamefont {L.}~\bibnamefont {Li}}, \bibinfo
  {author} {\bibfnamefont {Z.}~\bibnamefont {Li}}, \bibinfo {author}
  {\bibfnamefont {H.}~\bibnamefont {Ji}}, \bibinfo {author} {\bibfnamefont
  {A.}~\bibnamefont {Stern}}, \bibinfo {author} {\bibfnamefont
  {Y.}~\bibnamefont {Xia}}, \bibinfo {author} {\bibfnamefont {T.}~\bibnamefont
  {Cao}}, \bibinfo {author} {\bibfnamefont {W.}~\bibnamefont {Bao}}, \bibinfo
  {author} {\bibfnamefont {C.}~\bibnamefont {Wang}}, \bibinfo {author}
  {\bibfnamefont {Y.}~\bibnamefont {Wang}},  \emph {et~al.},\ }\href@noop {}
  {\bibfield  {journal} {\bibinfo  {journal} {Nature}\ }\textbf {\bibinfo
  {volume} {546}},\ \bibinfo {pages} {265} (\bibinfo {year}
  {2017})}\BibitemShut {NoStop}%
\bibitem [{\citenamefont {Xu}\ \emph {et~al.}(2018)\citenamefont {Xu},
  \citenamefont {Feng}, \citenamefont {Xiang},\ and\ \citenamefont
  {Bellaiche}}]{Xu2018}%
  \BibitemOpen
  \bibfield  {author} {\bibinfo {author} {\bibfnamefont {C.}~\bibnamefont
  {Xu}}, \bibinfo {author} {\bibfnamefont {J.}~\bibnamefont {Feng}}, \bibinfo
  {author} {\bibfnamefont {H.}~\bibnamefont {Xiang}}, \ and\ \bibinfo {author}
  {\bibfnamefont {L.}~\bibnamefont {Bellaiche}},\ }\href {\doibase
  10.1038/s41524-018-0115-6} {\bibfield  {journal} {\bibinfo  {journal} {npj
  Computational Materials}\ }\textbf {\bibinfo {volume} {4}} (\bibinfo {year}
  {2018}),\ 10.1038/s41524-018-0115-6}\BibitemShut {NoStop}%
\bibitem [{\citenamefont {Dai}\ \emph {et~al.}(2008)\citenamefont {Dai},
  \citenamefont {Xiang},\ and\ \citenamefont {Whangbo}}]{dai2008effects}%
  \BibitemOpen
  \bibfield  {author} {\bibinfo {author} {\bibfnamefont {D.}~\bibnamefont
  {Dai}}, \bibinfo {author} {\bibfnamefont {H.}~\bibnamefont {Xiang}}, \ and\
  \bibinfo {author} {\bibfnamefont {M.-H.}\ \bibnamefont {Whangbo}},\
  }\href@noop {} {\bibfield  {journal} {\bibinfo  {journal} {Journal of
  computational chemistry}\ }\textbf {\bibinfo {volume} {29}},\ \bibinfo
  {pages} {2187} (\bibinfo {year} {2008})}\BibitemShut {NoStop}%
\bibitem [{\citenamefont {Marzari}\ and\ \citenamefont
  {Vanderbilt}(1997)}]{Marzari1997}%
  \BibitemOpen
  \bibfield  {author} {\bibinfo {author} {\bibfnamefont {N.}~\bibnamefont
  {Marzari}}\ and\ \bibinfo {author} {\bibfnamefont {D.}~\bibnamefont
  {Vanderbilt}},\ }\href {\doibase 10.1103/PhysRevB.56.12847} {\bibfield
  {journal} {\bibinfo  {journal} {Phys. Rev. B}\ }\textbf {\bibinfo {volume}
  {56}},\ \bibinfo {pages} {12847} (\bibinfo {year} {1997})}\BibitemShut
  {NoStop}%
\bibitem [{\citenamefont {Kohn}(1959)}]{Kohn1959}%
  \BibitemOpen
  \bibfield  {author} {\bibinfo {author} {\bibfnamefont {W.}~\bibnamefont
  {Kohn}},\ }\href {\doibase 10.1103/PhysRev.115.809} {\bibfield  {journal}
  {\bibinfo  {journal} {Physical Review}\ }\textbf {\bibinfo {volume} {115}},\
  \bibinfo {pages} {809} (\bibinfo {year} {1959})}\BibitemShut {NoStop}%
\bibitem [{\citenamefont {Cheinet}\ \emph {et~al.}(2008)\citenamefont
  {Cheinet}, \citenamefont {Trotzky}, \citenamefont {Feld}, \citenamefont
  {Schnorrberger}, \citenamefont {Moreno-Cardoner}, \citenamefont {F\"olling},\
  and\ \citenamefont {Bloch}}]{Cheinet2008}%
  \BibitemOpen
  \bibfield  {author} {\bibinfo {author} {\bibfnamefont {P.}~\bibnamefont
  {Cheinet}}, \bibinfo {author} {\bibfnamefont {S.}~\bibnamefont {Trotzky}},
  \bibinfo {author} {\bibfnamefont {M.}~\bibnamefont {Feld}}, \bibinfo {author}
  {\bibfnamefont {U.}~\bibnamefont {Schnorrberger}}, \bibinfo {author}
  {\bibfnamefont {M.}~\bibnamefont {Moreno-Cardoner}}, \bibinfo {author}
  {\bibfnamefont {S.}~\bibnamefont {F\"olling}}, \ and\ \bibinfo {author}
  {\bibfnamefont {I.}~\bibnamefont {Bloch}},\ }\href {\doibase
  10.1103/PhysRevLett.101.090404} {\bibfield  {journal} {\bibinfo  {journal}
  {Phys. Rev. Lett.}\ }\textbf {\bibinfo {volume} {101}},\ \bibinfo {pages}
  {090404} (\bibinfo {year} {2008})}\BibitemShut {NoStop}%
\bibitem [{\citenamefont {Campbell}\ \emph {et~al.}(2006)\citenamefont
  {Campbell}, \citenamefont {Mun}, \citenamefont {Boyd}, \citenamefont
  {Medley}, \citenamefont {Leanhardt}, \citenamefont {Marcassa}, \citenamefont
  {Pritchard},\ and\ \citenamefont {Ketterle}}]{Campbell2006}%
  \BibitemOpen
  \bibfield  {author} {\bibinfo {author} {\bibfnamefont {G.~K.}\ \bibnamefont
  {Campbell}}, \bibinfo {author} {\bibfnamefont {J.}~\bibnamefont {Mun}},
  \bibinfo {author} {\bibfnamefont {M.}~\bibnamefont {Boyd}}, \bibinfo {author}
  {\bibfnamefont {P.}~\bibnamefont {Medley}}, \bibinfo {author} {\bibfnamefont
  {A.~E.}\ \bibnamefont {Leanhardt}}, \bibinfo {author} {\bibfnamefont {L.~G.}\
  \bibnamefont {Marcassa}}, \bibinfo {author} {\bibfnamefont {D.~E.}\
  \bibnamefont {Pritchard}}, \ and\ \bibinfo {author} {\bibfnamefont
  {W.}~\bibnamefont {Ketterle}},\ }\href@noop {} {\bibfield  {journal}
  {\bibinfo  {journal} {Science}\ }\textbf {\bibinfo {volume} {313}},\ \bibinfo
  {pages} {649} (\bibinfo {year} {2006})}\BibitemShut {NoStop}%
\bibitem [{\citenamefont {Stamper-Kurn}\ and\ \citenamefont
  {Ueda}(2013)}]{Stamper-Kurn13}%
  \BibitemOpen
  \bibfield  {author} {\bibinfo {author} {\bibfnamefont {D.~M.}\ \bibnamefont
  {Stamper-Kurn}}\ and\ \bibinfo {author} {\bibfnamefont {M.}~\bibnamefont
  {Ueda}},\ }\href@noop {} {\bibfield  {journal} {\bibinfo  {journal} {Reviews
  of Modern Physics}\ }\textbf {\bibinfo {volume} {85}},\ \bibinfo {pages}
  {1191} (\bibinfo {year} {2013})}\BibitemShut {NoStop}%
\bibitem [{\citenamefont {Hauschild}\ and\ \citenamefont
  {Pollmann}(2018)}]{Hauschild2018}%
  \BibitemOpen
  \bibfield  {author} {\bibinfo {author} {\bibfnamefont {J.}~\bibnamefont
  {Hauschild}}\ and\ \bibinfo {author} {\bibfnamefont {F.}~\bibnamefont
  {Pollmann}},\ }\href {\doibase 10.21468/SciPostPhysLectNotes.5} {\bibfield
  {journal} {\bibinfo  {journal} {SciPost Physics Lecture Notes}\ ,\ \bibinfo
  {pages} {5}} (\bibinfo {year} {2018})}\BibitemShut {NoStop}%
\bibitem [{\citenamefont {Vidal}(2004)}]{Vidal2004}%
  \BibitemOpen
  \bibfield  {author} {\bibinfo {author} {\bibfnamefont {G.}~\bibnamefont
  {Vidal}},\ }\href {\doibase 10.1103/PhysRevLett.93.040502} {\bibfield
  {journal} {\bibinfo  {journal} {Phys. Rev. Lett.}\ }\textbf {\bibinfo
  {volume} {93}},\ \bibinfo {pages} {040502} (\bibinfo {year}
  {2004})}\BibitemShut {NoStop}%
\end{thebibliography}%
